\documentclass[conference]{IEEEtran}
\IEEEoverridecommandlockouts
\usepackage{cite}
\usepackage{amsmath,amssymb,amsfonts}
\usepackage{algorithmicx}
\usepackage{graphicx}
\usepackage{textcomp}
\usepackage{xcolor}
\usepackage{stfloats}
\usepackage{algorithm}
\usepackage{algpseudocode}
\usepackage{breqn}

\def\BibTeX{{\rm B\kern-.05em{\sc i\kern-.025em b}\kern-.08em
    T\kern-.1667em\lower.7ex\hbox{E}\kern-.125emX}}

\IEEEoverridecommandlockouts \IEEEpubid{\makebox[\columnwidth]{ 979-8-3503-7128-4/24/$31.00$~\copyright2024 IEEE\hfill} \hspace{\columnsep}\makebox[\columnwidth]{ }} 

\begin{document}

\title{Multicore DRAM Bank-\& Row-Conflict Bomb for Timing Attacks in Mixed-Criticality Systems
}


 \author{\IEEEauthorblockN{Antonio Savino\textsuperscript{1}, Gautam Gala\textsuperscript{2}, Marcello Cinque\textsuperscript{1}, Gerhard Fohler\textsuperscript{2}}
 \IEEEauthorblockA{(1) Università degli Studi di Napoli Federico II, Italy \\
 (2) Technical University of Kaiserslautern-Landau, Germany\\
 anto.savino@studenti.unina.it, macinque@unina.it \{gala,fohler\}@eit.uni-kl.de}
}

\maketitle

\begin{abstract}
With the increasing use of multicore platforms to realize mixed-criticality systems, understanding the underlying shared resources, such as the memory hierarchy shared among cores, and achieving isolation between co-executing tasks running on the same platform with different criticality levels becomes relevant. 
In addition to safety considerations, a malicious entity can exploit shared resources to create timing attacks on critical applications. 
In this paper, we focus on understanding the shared DRAM dual in-line memory
module and created a timing attack, that we named the "bank \& row conflict bomb", to target a victim task in a multicore platform. 
We also created a "navigate" algorithm to understand how victim requests are managed by the Memory Controller and provide valuable inputs for designing the bank \& row conflict bomb. We performed experimental tests on a 2nd Gen Intel Xeon Processor with an 8GB DDR4-2666 DRAM module to show that such an attack can produce a significant increase in the execution time of the victim task by about 150\%, motivating the need for proper countermeasures to help ensure the safety and security of critical applications.
\end{abstract}


\section{Introduction}
Multicore processors improve performance while reducing SWaP (Size, Weight, and Power) compared to traditional single-core processors. As a result of these advantages and the anticipated mass market obsolescence for single-core processors, industries have already started using multicore processors in safety-critical domains such as automotive, and railway, which traditionally used only single-core processors. 

Achieving the required isolation in multicore processors is difficult as co-executing tasks on different processors can cause contention in the shared resources (such as shared bus, memory, and network). The contention in shared resources can lead to unpredictable delays and, in the worst case, cause deadline misses in safety-critical applications and reduce the quality of service (QoS) of non-critical applications. Often, temporal partitioning of cores and spatial partitioning of memory is used with meticulous planning and CPU scheduling to co-execute pre-planned mixed-criticality applications on multicore processors. Previous works have shown that despite these techniques, meeting deadlines is not always possible without considering shared resource regulation. 

The railway industry is exploring ways to benefit from cloud computing and entering a new market segment: safety-critical applications as cloud-based services \cite{RTcloud-ISORC2021}. The automotive industry is moving towards software-defined vehicles and is investigating Cloud-native as the design paradigm \cite{soafee}. Eker et al. \cite{ericsson} highlighted the need and requirements of real-time cloud computing in industry 4.0 use cases.

These upcoming safety-critical / real-time cloud use cases rely heavily on multicore processors for the underlying computing hardware and require open-world assumptions. We need to consider not only the co-execution of pre-planned mixed-criticality applications but also dynamically arriving applications, including malicious ones. Malicious entities can purposefully exacerbate the shared resource contention to cause targeted attacks on the timing behavior of safety-critical applications, causing deadline misses. Thus, carefully examining each shared resource in multicore processors is essential to discovering all possible interference channels. With this knowledge, it will be possible to adequately regulate them at runtime to provide timing guarantees for safety-critical applications despite unplanned or malicious applications.

Newer safety-critical and non-critical applications require large amounts of data and frequent memory accesses. A malicious entity can exploit the vulnerabilities in the shared memory hierarchy to create timing attacks on safety-critical applications. Some previous works have often focused on parts of memory hierarchy such as caches, shared buses, and memory controllers (e.g., \cite{7461353, giorgio, farinaISORC2022, monacoISORC2023, rdt, bank_coloring}). However, we took inspiration from the conceptual foundations laid by \cite{dos_analysis}\cite{dos_cpugpu}\cite{dos_mem_aware}. We focus on understanding commercial-of-the-shelf DDR4 DRAM dual in-line memory modules (DIMMs) and demonstrating how the DIMM architecture can be exploited to do a timing attack on a safety-critical application. 

First, we created a "navigate" algorithm that stresses the DRAM to help us understand how requests are managed and how parallelism is exploited to achieve higher performance. Based on understanding the DRAM module via the  "navigate" algorithm, we then show how a "row conflict bomb" can be created to cause timing attacks on a victim task. 
We performed experiments on a Dell R640 server with 2nd generation Intel Xeon Gold processor (Cascade Lake) \cite{intel} \cite{uncore} using an 8GB DDR4-2666 ECC RDIMM (1Rx8) DRAM module. However, the "navigate" algorithm can be used to understand other similar DDR4 DRAM DIMMs and create a "row conflict bomb" targeted to their architecture.

The following section clarifies some essential DRAM concepts. Section \ref{related_work} presents the related work. Section \ref{design} explains the design of the attack, and the two newly created algorithms (navigate and bank \& row conflict bomb). Section \ref{sec:exp} presents the experimental setup, the design of experiments, benchmarks, and discusses the results. Section \ref{sec:conclusion} provides the conclusion and the future work.

\vspace{-0.1cm}
\section{Background}
\begin{figure}[b]
    \centering
    \vspace{-0.6cm}
    \includegraphics[width=8cm]{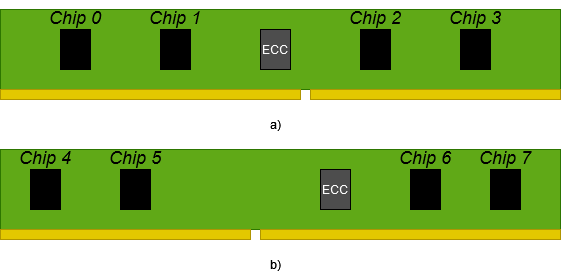}
    \vspace{-0.4cm}
    \caption{8GB DDR4-2666 ECC RDIMM (1Rx8): a) front side ; b) back side}
    \vspace{-0.4cm}
    \label{fig:dram}
\end{figure}

\begin{figure}[b]
    \centering
    \includegraphics[width=8cm]{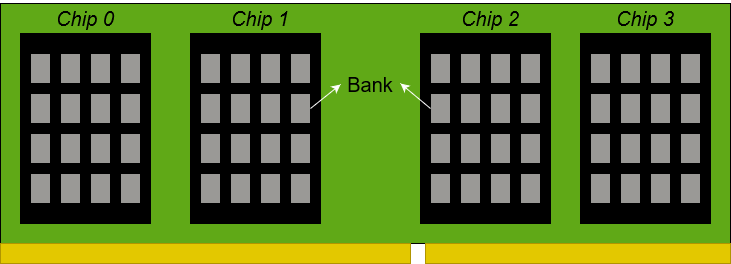}
    \vspace{-0.3cm}
    \caption{8GB DDR4-2666 ECC RDIMM (1Rx8): focus on the front side}
    \label{fig:front}
\end{figure}

Modern Dynamic Random Access Memory modules, as shown in Figures~\ref{fig:dram},~\ref{fig:front}), are structured in a complex way, with different levels like channels, ranks, bank groups, and banks. These components work together to store and retrieve data efficiently.
Inside a multicore CPU, there is an integrated memory controller responsible for managing the DRAM modules. Sometimes, there can be more than one memory controller within a single CPU, each of which can handle multiple DRAM channels. 
They work independently and can handle data at the same time, which makes memory operations faster.
Memory modules, known as DIMMs (Dual In-line Memory Modules), typically have one or two ranks, which are like sections of the module. Each rank contains multiple banks, and several banks can be used at the same time. DDR4 RAM, a common type of memory, has 16 banks organized into four groups, with four banks in each group.

        
Inside these banks are rows of memory cells, which can be activated together. But one can only access them through a row buffer, a temporary storage place for one row of data. The row buffer acts like a special kind of cache. When reading or writing data from a DRAM row, the first step is to load the data into the row buffer using an Activate (ACT) command. Afterward, the data can be manipulated using a CAS command. To access a different row within the same bank, returning the row buffer to the memory array with a Pre-charge (PRE) command is necessary before activating the new row. Because of how DRAM works, it needs to be periodically refreshed to maintain data integrity. This involves writing back all the row buffers to the memory array, which means they become empty after a refresh.
In summary, during the memory access process, the memory controller picks a channel based on the address, and each channel has its own set of addresses and data lines. Ranks and banks within a channel share these lines, so extra bits are used to select the right one. The mapping function determines the activation of the rank and bank associated with accessing data at a designated memory address. However, there are some timing constraints and challenges in efficiently managing these memory operations, especially when multiple applications run on different CPU cores and access the same memory bank \cite{dos_mem_aware}. This can lead to slower performance and complex timing issues. To solve some of these problems \cite{bank_coloring} \cite{tintmalloc} are proposals that use a private bank scheme, where each CPU core gets its own set of memory banks. However, these approaches come with challenges, such as needing changes to the memory controllers and being less flexible. Hence, they do not apply to commercial-off-the-shelf (COTS) multicore processors.

\begin{figure}
    \centering
    \includegraphics[width=8cm]{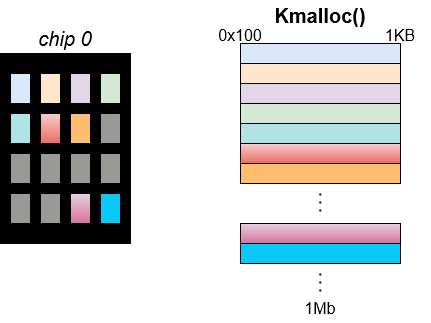}
    \vspace{-0.4cm}
    \caption{Request handling by a DRAM chip with high-order interleaving}
    \vspace{-0.4cm}
    \label{fig:parallel_access}
\end{figure}

\paragraph*{\textbf{Interleaving}}
Interleaving is a technique used in DDR4 RAM to enhance performance. Instead of storing consecutive data within the same memory block, interleaving spreads it across different blocks, Figure~\ref{fig:parallel_access}.
This approach enables parallel access to data and can improve performance, especially with a wide data bus. Additionally, interleaving can be managed by motherboards and configured through the BIOS. 
It is important to note that various interleaving modes exist, such as high-order and low-order interleaving.

\begin{figure}
    \centering
    \includegraphics[width=8cm]{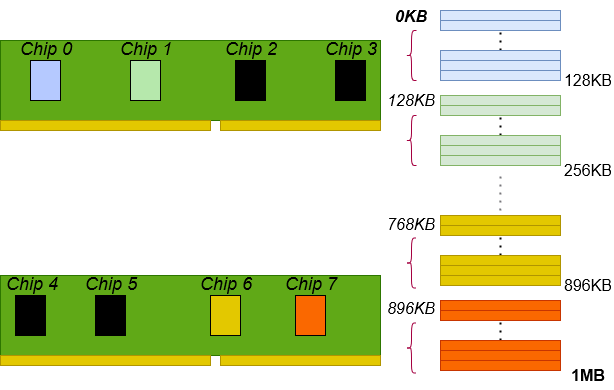}
    \vspace{-0.5cm}
    \caption{High-order interleaving}
    \vspace{-0.5cm}
    \label{fig:ho_interleaving}
\end{figure}
In high-order interleaving (Figure~\ref{fig:ho_interleaving}),  also known as Memory Banking, the most significant bits of the address select the memory chip. The least significant bits are sent as addresses to each chip. A problem could be that consecutive addresses tend to be in the same chip; this is one of the vulnerabilities that will be exploited in Section \ref{row_bomb} on the target architecture we experimented with.
Conversely, in low-order interleaving, the least significant bits select the memory bank. Consecutive memory addresses are in different memory modules. 

Using interleaving can enhance efficiency in memory access, reducing delays and making better use of memory bandwidth. However, the choice of interleaving mode depends on the application's specific needs and workload.


\section{Related Work}\label{related_work}
Nowadays, in embedded systems, heavy workloads involving a large amount of data are increasingly prevalent, characterized by frequent memory accesses. This trend is particularly evident in critical real-time systems, which have the prerogative of working with substantial datasets that require significant computing power, e.g., in vision-based control of autonomous micro-air vehicles \cite{vision-based}. Such \textit{critical tasks} have a high volume of data to be handled and require frequent memory accesses for which they compete with normal tasks.

Managing a mixed-criticality system poses a challenge:\textit{\textbf{ shared resources}}. Different tasks access these resources, and if not regulated and standardized, they could be considered real \textit{vulnerabilities} that third parties could exploit to create an attack undermining a system's dependability, reliability, security, and safety.

\subsection{Shared Resource Vulnerability}\label{shared}
We drew deep inspiration from the investigations conducted in previous works, such as \cite{dos_analysis},\cite{dos_cpugpu},\cite{dos_mem_aware}. These authors established a robust conceptual foundation, focusing on the centrality of our attack. They comprehensively outlined the vulnerabilities arising from targeting DRAM in architecture, providing details on exploiting row conflicts to increase the Worst-Case Execution Time (WCET). 
Despite the different methodologies, our work differs in the approach to the problem, basing the centrality in understanding how requests are handled by the DRAM and how the latter manages resources a pool of physically contiguous requests. In fact, the work done by the Navigate Algorithm in the \ref{navigate} Section helps to understand how DRAM exploits parallelism in order to create an algorithm (Section \ref{row_bomb}), which stresses even more the DRAM based on the data obtained.

Although with a slightly different focus, relevant contributions were provided by \cite{ farinaISORC2022, monacoISORC2023, rdt}. While not presenting a direct attack, the works focus on a workload that stresses the memory within the considered architecture. The main goal of this study is to measure memory performance, offering valuable insights that, though not aligned with the primary purpose of our work, contribute to the overall understanding of memory contention impact in a similar context.

The integration of these approaches was accomplished with the contribution of \cite{parallelism_aware},\cite{permutation_based},\cite{DRAMA},\cite{taming}. 
These approaches extensively examined the various components used in our paper as potential vulnerabilities for an attack. These thorough analyses proved crucial in providing a comprehensive overview of potential weaknesses in the architecture, forming a solid foundation for our work.

\subsection{Reverse Engineering}\label{reverse}
In the increasingly intricate landscape of mixed-criticality systems, more and more works are emerging focusing on identifying data mapping in memory and understanding the modes of accessing such data. However, it is crucial to note that these works generally make assumptions that are often limiting and require a deep understanding of how data is mapped in memory. As seen in Section \ref{shared}, these assumptions can also extend to a targeted use of virtual memory; this could be motivated by various factors, such as performance enhancement, TLB reduction, or even cache coherence improvement.

An example of research focused on data mapping in memory is PALLOC \cite{palloc}, a tool to conduct in-depth reverse engineering of mapping bits. PALLOC is a kernel-level memory allocator that leverages virtual-to-physical memory translation based on pages to selectively allocate memory pages of each application to the desired DRAM memory banks. However, this presupposes that such a patch is implemented in the kernel of the target architecture, making it highly unlikely to stage an attack that undermines the security of an architecture introduced into a real-world scenario.
The creation of such tools has been instrumental in bridging gaps in understanding data mapping mechanisms, enabling industry experts to explore the memory structure more thoroughly.

Alternative approaches adopt a \textit{timing-based} perspective, founded on the principle that a row buffer hit results in lower access latency than a row buffer conflict. This concept, also reflected in our work, is also used by \cite{DRAMA}. Specifically, they focus on finding a linear function through a brute-force method to understand memory mapping. The analysis of mapping bits, implemented in this context, provides a clear overview of the arrangement of data in memory and how they are handled during the execution of various tasks. On the other hand, however, there is the complexity in obtaining this information; usually, the architectures do not provide adequate documentation, which increases the probability of error in reaching such data, especially with these tools, which give a probabilistic and non-deterministic indication of the mapping function.
Another strategy to determine data mapping, avoiding the timing-based approach, is using PMC (Performance Monitoring Counters), as illustrated in \cite{reverse_eng_pmc}, specifically focusing on an architectural method for Intel platforms. This approach stands out for its potential to overcome limitations associated with timing-based approaches. Following a timing-based approach in our work is challenging due to many issues, such as the potential delays introduced by processing other instructions in the pipeline. Additionally, the memory controller represents another source of possible inaccuracies, as it can reorganize DRAM access requests, altering the timing and influencing row buffer access behavior. In general, using PMC is a promising strategy to overcome the limitations of timing-based approaches and contribute to a more accurate understanding of data mapping in mixed-criticality systems.

The complexity of obtaining detailed information and the lack of adequate documentation in architectures can make identifying data mapping more challenging and error-prone. However, despite these challenges, using advanced tools, such as those mentioned, can still represent a crucial step forward in overcoming limitations and gaining a deeper understanding of memory dynamics in mixed-criticality systems.

\subsection{Memory Contention Mitigation}
Many works specifically address the regulation (e.g., memory bandwidth allocation) and monitoring (e.g., number of memory accesses) of a system's resources to achieve temporal isolation between applications. They identify elements in an architecture (e.g., PMC) that can be exploited for monitoring contention in shared resources. The hardware-based approach of Process Cruise Control \cite{cruise_control} dynamically allows throttling processes that exceed a certain number of memory operations within a specific interval or using the number of TLB-miss handlers to employ a certain number of NOPs to reduce the number of memory accesses by a thread. If software-based throttling mechanisms are desired, a periodic budget can be given to a thread, limiting the number of memory accesses from the thread when the budget is exhausted for the period (similar to an aperiodic server with CPU bandwidth reservation) \cite{mem_access}. Memguard \cite{memguard} attempts to regulate memory bandwidth allocation for individual cores to control memory accesses and improve temporal isolation between cores. These methods do not optimally exploit memory throughput, often keeping lower-criticality/best-effort processes stalled for longer than necessary and reducing system performance.

There also exist hardware technologies \cite{rdt}, such as the Intel Resource Director Technology (RDT), used to regulate contention in shared resources. Intel RDT has two relevant capabilities: Cache Allocation Technology (CAT) and Memory Bandwidth Allocation (MBA). CAT implements cache way-partitioning at the hardware level, usually of the Last Level Cache \cite{holistic}\cite{vcat}\cite{dna} but is not very efficient against temporal isolation, as demonstrated in \cite{llc_intel}\cite{dos_cpugpu}. MBA delays the requests going to the interconnect from a core's private context. Through ten configurable delay values, we can apply indirect memory bandwidth throttling. In its short life, it has seen different versions, initially explored in works such as \cite{farinaISORC2022}\cite{copart}\cite{emba}. However, due to many bugs and limitations, Intel RDT did not yield the expected results for use in real-time systems, as shown in work \cite{giorgio}. Farina et al. \cite{giorgio, farinaISORC2022} demonstrated substantial variability in the results obtained and how MBA is decisive only when set beyond a particular delay value. Recently, Monaco et al. \cite{monacoISORC2023} explored combining low overhead, imprecise hardware regulation technology (e.g., Intel's RDT) together with high overhead, strict software memory regulation techniques (based on the number of memory accesses per application) to reap benefits of both approaches.


\section{Design of Attack}\label{design}

\begin{figure}
    \centering
    \includegraphics[width=8cm]{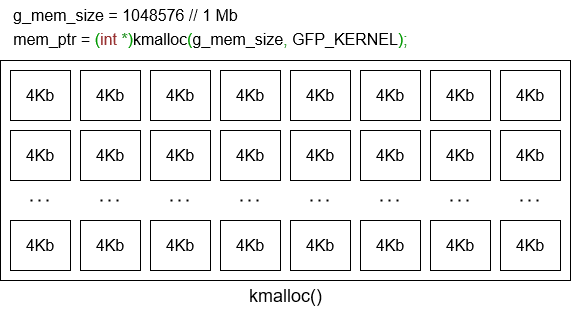}
    \vspace{-0.4cm}
    \caption{Kmalloc()}\vspace{-0.2cm}
    \label{fig:kmalloc}\vspace{-0.4cm}
\end{figure}

To implement our algorithm, it is essential to elucidate the underlying assumptions and the context in which the entire experiment unfolds. It is significant to specify that our experiment is embedded within a kernel module, a detail reiterated both in the explanation of the Navigate algorithm in Section \ref{navigate} and in the explanation of the DRAM Bank \& Row-Conflict Bomb in Section \ref{row_bomb}. The module employs a function named kmalloc(); this function strongly resembles its counterpart in user space, \textit{malloc()}, with the distinction that flags can be added here for enhanced memory allocation management. Kmalloc() returns a pointer to a memory area of at least the requested size in length. It is important to note that the allocated \textit{memory block is physically contiguous} - a characteristic to be leveraged for comprehending how DRAM Chips handle a pool of physically contiguous requests. The memory block is structured as follows in Figure~\ref{fig:kmalloc}.
\begin{table}[b]
    \vspace{-0.4cm}
    \centering
    \caption{Total size of a single DRAM bank}
    \vspace{-0.2cm}
    \label{table:1}
    \begin{tabular}{||p{2cm} p{3cm} p{2.5cm}||} 
     \hline
     Row Address Bit & Column Address Bit & Width of each column\\ [1ex] 
     \hline\hline
        \multicolumn{1}{|c|}{from $A_0$ to $A_{15}$} & \multicolumn{1}{|c|}{from $A_0$ to $A_{9}$} & \multicolumn{1}{|c||}{} \\ 
        \multicolumn{1}{|c|}{$2^{16}$} & \multicolumn{1}{|c|}{$2^{10}$} & \multicolumn{1}{|c||}{$2^{3}$} \\
        \multicolumn{1}{|c|}{65,536 bit} & \multicolumn{1}{|c|}{1,024 bit} & \multicolumn{1}{|c||}{8 bit} \\
        \hline
        \hline
       \multicolumn{3}{||c||}{{\fontsize{8}{10}\selectfont 65,536 * 1,024 * 8 = 536,870,912 bit}} \\ [1ex] 
        \hline
    \end{tabular}\vspace{0.1cm}
    
    \centering
    \caption{DRAM chip capacity calculation}\label{tab:chip}
    \vspace{-0.2cm}
    \begin{tabular}{|l|l|l|}
    \hline
        \# of Bank Groups & 4 & ~ \\ \hline
        \# of Bank per Group & 4 & ~ \\ \hline
        Total Number of Banks & 16 & ~ \\ \hline
        \hline 
        & \textbf{bit} & \textbf{GB} \\ \hline
        BankSize*NumBanks & 8.589.934.592 & 1 \\ \hline
    \end{tabular}\vspace{0.1cm}
    
    \centering
    \caption{DRAM capacity calculation}
    \label{tab:dram}
    \vspace{-0.2cm}
    \begin{tabular}{|l|l|l|}
    \hline
        Number of chips & 8 & ~ \\ \hline
        \hline 
        & \textbf{bit} & \textbf{GB} \\ \hline
        \textbf{DRAM Capacity} & 68.719.476.736 & 8 \\ \hline
    \end{tabular}
\end{table}

To calculate the capacity of the DRAM, we checked the datasheet of the DRAM installed in the target architecture (Tables~\ref{table:1}, \ref{tab:chip}, and \ref{tab:dram}). We determined how many bits are allocated for the number of rows and columns and the width of each column (Table \ref{table:1}). Table \ref{table:1} shows that the number of columns in a DRAM bank is 1KB. Upon surpassing this size, a new row will be \textit{opened}, and the row buffer will become marked as \textit{dirty}. We will utilize this information in the following Sections.

\subsection{Threat Model}\label{threat_model}
In designing the threat model for the implementation of the experiment it is crucial to consider potential vulnerabilities and security risks inherent in the system. The following threat model outlines key assumptions, context, and potential threats:
\begin{itemize}
      \item \textit{\textbf{Assumptions 1} - Kernel Module Environment}: The experiment operates within a kernel module, leveraging the kmalloc() function for memory allocation management.
            
      \item \textit{\textbf{Assumption 2} - Physical Contiguity of Allocated Memory}: Memory blocks allocated by kmalloc() are physically contiguous, which impacts the understanding of how DRAM chips handle requests.

      \item \textit{\textbf{Context}}: The threat model focuses on the security implications within the kernel module, specifically concerning memory allocation and DRAM chip behavior.

      \item \textit{\textbf{Potential Threats} - Denial-of-Service (DoS)}: Malicious actors could potentially exploit the behavior of DRAM chips, causing excessive row openings or dirtying of row buffers, leading to resource exhaustion and system instability.

      \item \textit{\textbf{Mitigation Strategies}}: Enforce limits on monitoring DRAM chip behavior to mitigate potential DoS attacks.
\end{itemize}

\vspace{-0.25cm}
\subsection{Navigate Algorithm}\label{navigate}
We created the \textit{Navigate algorithm} to stress the DRAM and help us understand 1) how requests are managed, 2) how DRAM exploits parallelism, and 3) how DRAM chips handle a pool of physically contiguous requests. Before exploring the workflow of this algorithm, attention is directed towards the anticipated outcomes of the experiment. With the algorithm situated in a Linux kernel module, a contiguous memory block is allocated using kmalloc(), and distinct cores are assigned to a victim and malicious task(s). This setup enables the examination of individual core performance without interference. Observing the clock\_cycle of the victim task is crucial to understanding its benchmark completion, which the attackers may negatively impact. While the victim consistently accesses the same initial position in the kmalloc() block, the attacker threads traverse the entire memory block in successive iterations. The objective is to retrospectively analyze how DRAM chips handled these requests and comprehend how the memory controller sorted requests from a contiguous memory block. This emphasis is crucial, as this analysis will precisely elucidate how each chip manages contiguous requests. This insight is used in the \ref{row_bomb} section to determine the number of attacker threads and the offset, influencing competition for the same memory bank where the victim task is situated.


\begin{algorithm}
\caption{MSR Driver}\label{code:msr}
\begin{algorithmic} 
\State $zalloc\_cpumask\_var(\&cpu\_mask\_all, GFP\_NOWAIT)$
\State $global\_point\_state\_cpu \leftarrow alloc\_percpu(struct\ state\_cpu)$
\State $benchmark\_setup()$
\For{$i \leftarrow 0$ TO $num\_core\_enable$}
    \State $cpumask\_set\_cpu(i, cpu\_mask\_all)$ 
\EndFor
\State $cpus\_read\_lock()$
\ForAll{$id$ \textbf{in} $cpu\_mask\_all$}
    \State $state\_core \leftarrow per\_cpu\_ptr(global\_point\_state\_cpu, id)$
    \State $memset(state\_core, 0, sizeof(struct\ state\_cpu))$
\EndFor
\State $cpus\_read\_unlock()$
\State $on\_each\_cpu\_mask(cpu\_mask\_all, \textit{\textbf{benchmark}} , NULL, 1)$
\end{algorithmic}
\end{algorithm}

\begin{figure}
    \centering
    \vspace{-0.5cm}
    \includegraphics[width=6cm]{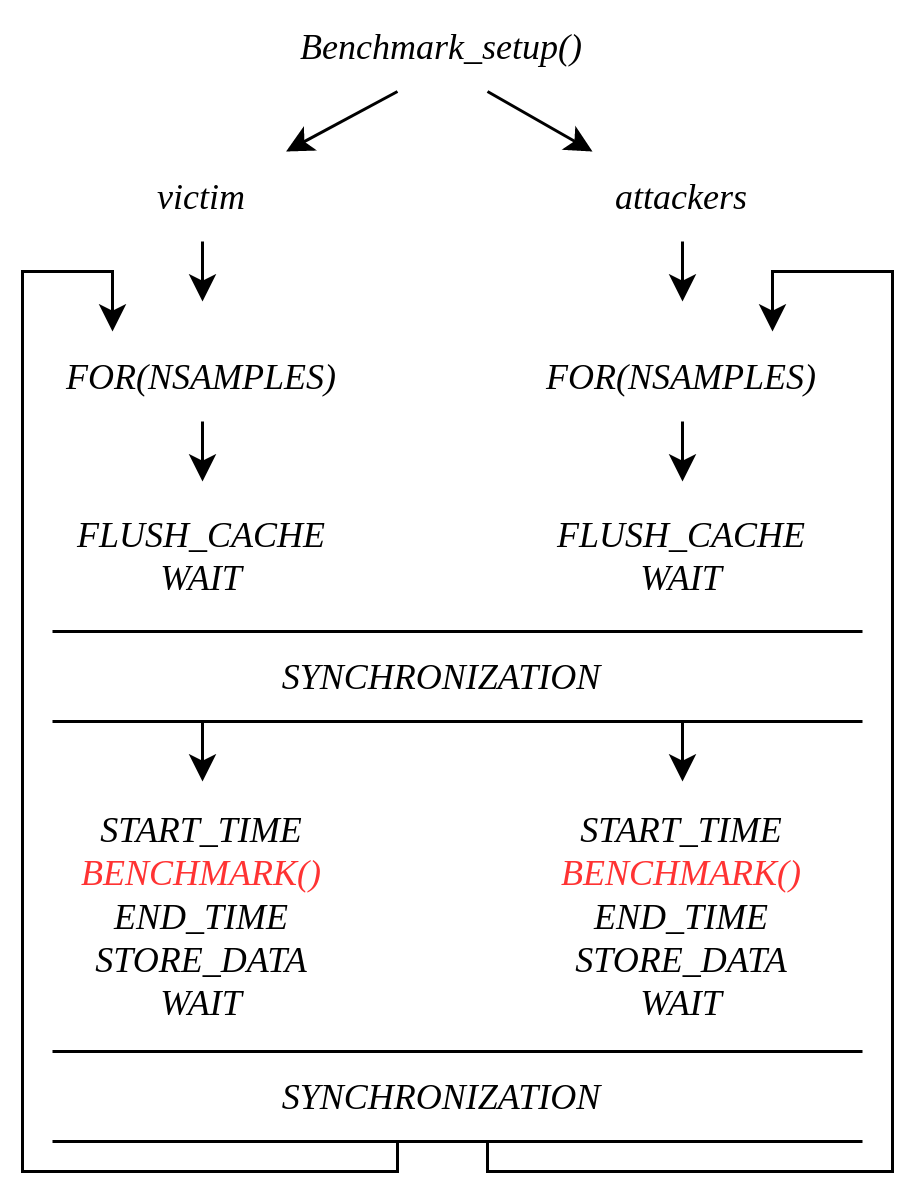}
    \vspace{-0.4cm}
    \caption{Workflow}
    \vspace{-0.6cm}
    \label{fig:workflow}
\end{figure}
\begin{figure}[h]
  \centering
  \begin{minipage}{0.6\linewidth}
    \includegraphics[width=\linewidth]{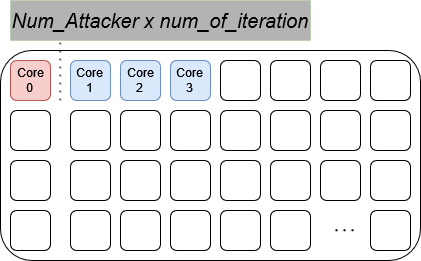}
  \end{minipage}%
  \vspace{0.1cm}
  \hfill
  \begin{minipage}{0.6\linewidth}
    \includegraphics[width=\linewidth]{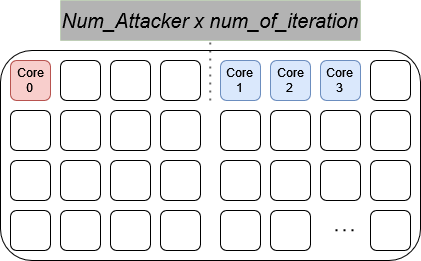}
  \end{minipage}%
  \vspace{0.1cm}
  \hfill
  \begin{minipage}{0.6\linewidth}
    \includegraphics[width=\linewidth]{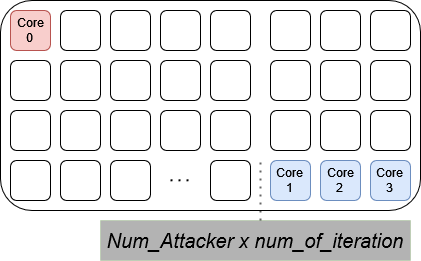}
  \end{minipage}
  \caption{Workflow logic}\vspace{-0.5cm}
  \label{fig:rationale}
\end{figure}

Figures~\ref{fig:workflow},~\ref{fig:rationale} and Algorithm~\ref{code:msr} depict the functioning of the code. The first step is to start by explaining the workflow of the entire algorithm and how it synchronizes the victim and attacker to ensure coherence when launching a single experiment.
The benchmarks performed by the victim and the attacker start at the exact same moment. This is crucial and fundamental for a better understanding of DRAM behavior in handling requests across different chips and banks.

\begin{enumerate}
    \item \textit{Initialization}: After mounting the kernel module, the algorithm begins by initializing the memory blocks, allocating space using kmalloc(), and configuring the victim and attacker processes.

    \item \textit{Contending for Memory}: The attacker and victim processes start contending for memory resources. They generate memory access requests in parallel.

    \item \textit{Synchronization}: The algorithm employs synchronization mechanisms to ensure synchronization. This might involve using signals, semaphores, or other synchronization primitives to coordinate the timing of memory accesses between the attacker and victim. This synchronization ensures that both processes start their operations simultaneously.

    \item \textit{Data Collection}: As the attacker and victim task access memory, the algorithm collects the clock\_cycles used for each victim and attacker benchmark until the entire workload is terminated.
    
    \item \textit{Iteration and Experimentation}: The algorithm likely repeats this process for a specified number of iterations or experiments. This helps gather a more comprehensive dataset for analysis and also reliable data.
\end{enumerate}
\begin{figure*}[t]
    \centering
    \includegraphics[width=12cm]{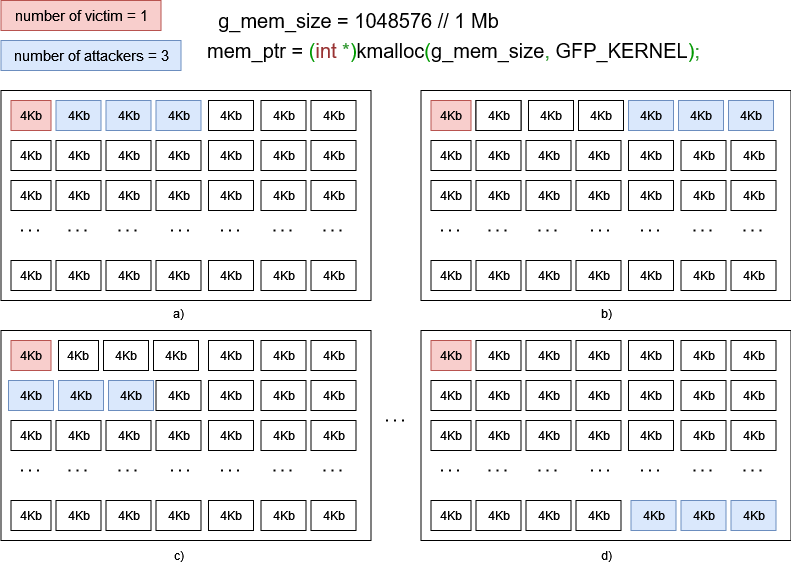}
    \vspace{-0.4cm}
    \caption{a) First iteration;  b) Second iteration;   c) Third iteration;   d) 85th iteration.}
    \vspace{-0.4cm}
    \label{fig:kmalloc_navigate}
\end{figure*}

As previously mentioned, this algorithm traverses the entire 1 MB memory block created using kmalloc() (Figure~\ref{fig:kmalloc_navigate}). For example, considering that 1 MB is fully traversed in 256 iterations (reads/writes of 4 KB each), in this case, the number of samples, denoted with NSAMPLES, depends on the chosen number of attacker threads.

Every single benchmark runs on a different core. Each attacker knows what his core number is (using \textit{id\_cpu = smp\_processor\_id()}).
We will require (256-1)/5 = 51 NSAMPLES if we have five attacker threads. Similarly, if we choose nine attacker threads, we will require (256-1)/9 = 28.3 = 28 NSAMPLES. \textit{We cannot consider all 256 NSamples because we must also consider the victim task space.}
Figure~\ref{fig:kmalloc_navigate} is an example with three attacker threads (totaling 85 NSAMPLES). Starting from this consideration, we can use this line of code before running the benchmark:
\begin{dmath*}
    index = id\_cpu + num\_attacker\_threads
    *num\_of\_iteration;
\end{dmath*}

\subsection{DRAM Bank \& Row-Conflict Bomb}\label{row_bomb}
Following the deployment of the Navigate Algorithm and subsequent analysis of clock\_cycle data, a crucial piece of information has been obtained: \textit{\textbf{we know the number of requests fulfilled by a single chip}} before the memory controller assigns another chip to address the remaining requests. Armed with this information, coupled with the details provided in Table \ref{tab:chip} and Table \ref{tab:dram}, indicating 16 DRAM banks per chip, the idea was to devise an algorithm much more effective than a mere contention within the chips. The purpose of the DRAM Bank \& Row-Conflict Bomb is to direct contention within a specific DRAM bank and then observe the victim's clock\_cycles to assess whether we have increased the Worst-Case Execution Time (WCET) of the entire workload.

\begin{figure}
    \centering
    \includegraphics[width=8cm]{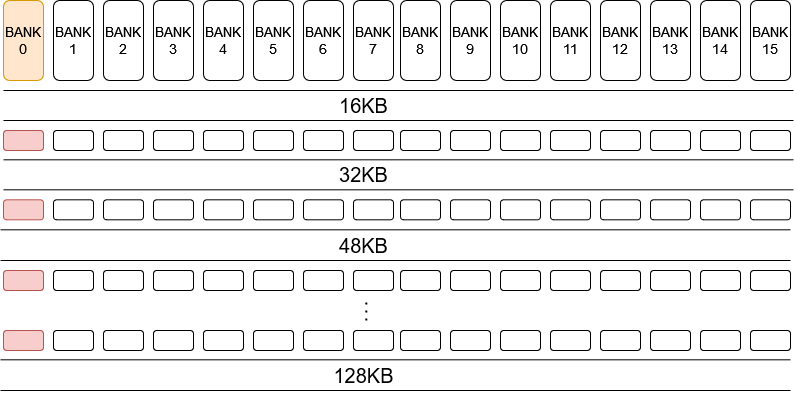}
    \vspace{-0.4cm}
    \caption{Benchmark: 1KB with 1 victim and 7 attacker threads}
    \vspace{-0.4cm}
    \label{fig:Row_Algorithm}
\end{figure}

The DRAM Bank \& Row-Conflict Bomb, in contrast to the Navigate algorithm, operates in a straightforward manner. In this scenario, the victim activity consistently accesses the initial memory location. At the same time, a predetermined number of attacker threads exclusively contend within the same DRAM bank by leveraging the offset, as shown in Figure \ref{fig:Row_Algorithm}. When the victim task accesses the first DRAM bank, the attacker threads are strategically deployed so that each lands in the same DRAM bank as the victim, thereby avoiding unnecessary contention across other DRAM banks within the chip. By jumping 15 banks at a time, the attacker threads consistently access the same DRAM bank as the victim. Navigating through the entire megabyte created with kmalloc() is not a concern, as it would render the attack illogical. 

The default number of attacker threads is set at 7, considering that, as depicted in Figure~\ref{fig:Row_Algorithm}, and with each chip handling one-eighth of the MB at a time (equivalent to 128 KB), 7 attacker threads are chosen to create contention within the same chip.
 From this algorithm, we expect worse results to reach the WCET for the victim task, as the contention of the row buffer in the DRAM must be tested due to the continuous row conflicts that are created. 

\section{Experimental Results}\label{sec:exp}
The characteristics of the architecture used for the experiment are outlined below to provide data that can be utilized in the future. In this case, it involves the 2nd Generation Intel Xeon Processor Scalable Family based on Cascade Lake product \cite{intel} \cite{uncore} with the Intel Xeon Gold 5218 CPU @ 2.30GHz processor with 8GB DDR4-2666 ECC RDIMM (1Rx8) DRAM module.
Another characteristic worth mentioning is the serial number of the DRAM installed (HMA81GR7CJR8N) in the architecture. It's important to note that, to stress the architecture with relatively small benchmarks, only one DRAM module has been installed, leaving the other slots empty.

\subsection{Design of Experiment: Navigate Algorithm}
We set up four different types of tests to achieve the algorithm's goals. The algorithm involves reading and writing data, so we first focus on one type and then switch to another. 

The \textit{goal} of our experiment is \textit{to stress} DRAM, allowing us to observe a significant increase in a specific memory bank's performance. Additionally, we have the opportunity to understand how chips in a DRAM module work together. Through the experiment, we also analyze how much memory a DRAM chip manages and if there is any cyclic behavior.

There are two main \textit{Factors} in our experiment: 1) \textit{number of cores}, i.e., the number of attacker threads, and 2) the \textit{size of the attacker and victim benchmark}. A benchmark consists of either consecutive memory writes (write benchmark) or memory reads (read benchmark), and these writes/reads can have a size of 1 KB, 4 KB, 8 KB, or 16 KB. Changing these two factors affects the stress on the DRAM.

\begin{table}[!ht]
    \centering
    \caption{Design of experiment - writing}\vspace{-0.2cm}
    \begin{tabular}{|p{3cm}|p{4cm}|}
    \hline
        \textbf{\textit{Goal}} & Study the trend of the clock\_cycles in order to see contention in the DRAM \\ \hline
        \hline
        \textbf{\textit{Factors}} & ~ \\ \hline
        number of cores & $n \in \{2,\ldots,11\}$ \\ \hline
        type of benchmark &$b \in \{b_1,b_2,b_3,b_4\}$ \\ \hline
        \hline
    \end{tabular}
    \vspace{-0.4cm}
\end{table}



\begin{table}[!ht]
    \centering
    \caption{Design of experiment - reading}\vspace{-0.2cm}
    \begin{tabular}{|p{3cm}|p{4cm}|}
    \hline
        \textbf{\textit{Goal}} & Study the trend of the clock\_cycles to observe contention in the DRAM \\ \hline
        \hline
        \textbf{\textit{Factors}} & ~ \\ \hline
        Number of cores & $n \in \{2,\ldots,11\}$ \\ \hline
        type of benchmark & $b' \in \{b'_1, b'_2, b'_3, b'_4\}$ \\ \hline
        
        \hline
    \end{tabular}
    \vspace{-0.4cm}
\end{table}

\subsection{Discussion Data: Navigate Algorithm}
We focus on the \textit{write benchmark} to simplify the content and avoid redundancies in visual explanations. Similar results were observed for the \textit{read benchmark}.
The conducted experiments are summarized in Table \ref{tab:Experiment}.

\begin{table}[h]
\centering
\caption{Total number of experiments conducted for various size benchmarks and workloads}\vspace{-0.2cm}
\begin{tabular}{|c|c|c|}
\hline
\textbf{Size Benchmark} & \textbf{Size Workload} & \textbf{Num Samples} \\
\hline
\hline
\textbf{\textit{1KB}} & \begin{tabular}[c]{@{}c@{}}256 KB \\ 512 KB\end{tabular} & \begin{tabular}[c]{@{}c@{}}256 x 1000\\ 512 x 1000\end{tabular} \\
\hline
\hline
\textbf{\textit{4KB}} & \begin{tabular}[c]{@{}c@{}}1024 KB \\ 2048 KB\end{tabular} & \begin{tabular}[c]{@{}c@{}}256 x 1000\\ 512 x 1000\end{tabular} \\
\hline
\hline
\textbf{\textit{8KB}} & \begin{tabular}[c]{@{}c@{}}2048 KB \\ 4096 KB\end{tabular} & \begin{tabular}[c]{@{}c@{}}256 x 1000\\ 512 x 1000\end{tabular} \\
\hline
\hline
\textbf{\textit{16KB}} & \begin{tabular}[c]{@{}c@{}}2048 KB \\ 4096 KB\end{tabular} & \begin{tabular}[c]{@{}c@{}}128 x 1000\\ 256 x 1000\end{tabular} \\
\hline
\end{tabular}
\label{tab:Experiment}
\vspace{-0.4cm}
\end{table}

For each benchmark size, two workload sizes were run. This was done to highlight through the collected clock\_cycles how those of the victim changed when the chip in which it was located was reopened by the attacker threads.

Within the first 15 memory locations distant from where the victim accesses memory, 15 writes do not significantly impact the number of victim clock cycles. However, a change occurs at the 16th location.

\begin{figure*}[bt]
    \centering
    \includegraphics[width=17cm]{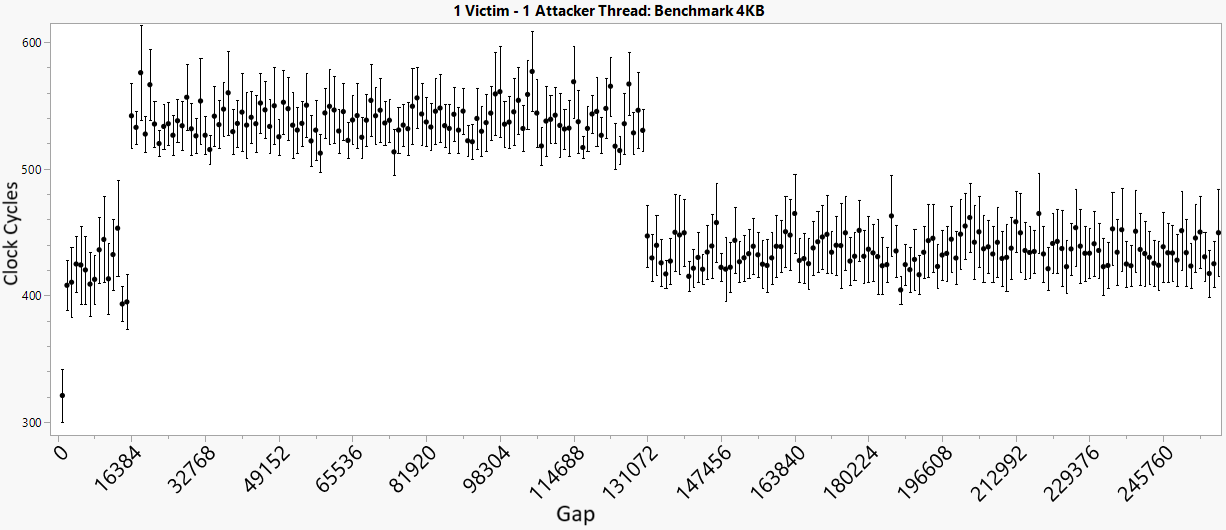}
    \caption{Workload 256 KB / benchmark 1KB: 1 victim \& 1 attacker}
    \label{fig:navigate_bomb}
\end{figure*}

Referring to the table \ref{tab:Experiment}, in this initial experiment, a total workload of 256KB is used, with each individual benchmark being 1 KB. Each request made by each attacker fills an entire DRAM page row (Figure~\ref{fig:navigate_bomb}). Considering the interleaving of DRAM, wherein a contiguous portion of memory is managed across multiple DRAM banks and chips to exploit DRAM parallelism, with our hardware, an 8GB DDR4-2666 ECC RDIMM (1Rx8) DRAM module, we have 16 DRAM banks. These banks get accessed in the first 15 iterations, causing every row buffer in the chip to be "open." However, from the 16th iteration, the attacker will force all row buffers to change, which continues as long as it remains within the victim's chip.  The contention sought for implementing a Denial Of Service with the DRAM Bank \& Row-Conflict Bomb is precisely observed at this point. The victim's clock cycles must be scrutinized, noting their increase to demonstrate the desired result.

\textit{How many reads must the attacker perform to ensure they always stay within the victim's chip before being distributed to other chips? }

Based on various tests, it was observed that this architecture manages 1MB of memory using all eight chips. Converting these numbers to bytes, 1048576 bytes (1MB) divided by the number of chips (8) equals 131072 bytes (128KB). This is precisely the number on the plot's X-axis where contention ends within the same DRAM chip. 

This is a detail not to be underestimated as our algorithm showed (in Figure~\ref{fig:navigate_bomb}) how many consecutive requests are executed in the same chip without any assumptions made on the operating system code and without any type of mapping as done in other works previously mentioned in Section \ref{related_work}. The DRAM Bank \& Row-Conflict Bomb explained in Section \ref{row_bomb}, uses precisely this data to calculate how many attacker threads will be chosen to create contention in that chip and in that specific DRAM bank.
Returning to the discussion of the data obtained through the Navigate Algorithm, the attacker will access other chips from now until the end of the number of samples. At the same time, the victim remains in chip 0. This no longer creates contention, and this behavior is visible until the end of the plot.

\begin{figure*}[bt]
    \centering
    \includegraphics[width=17cm]{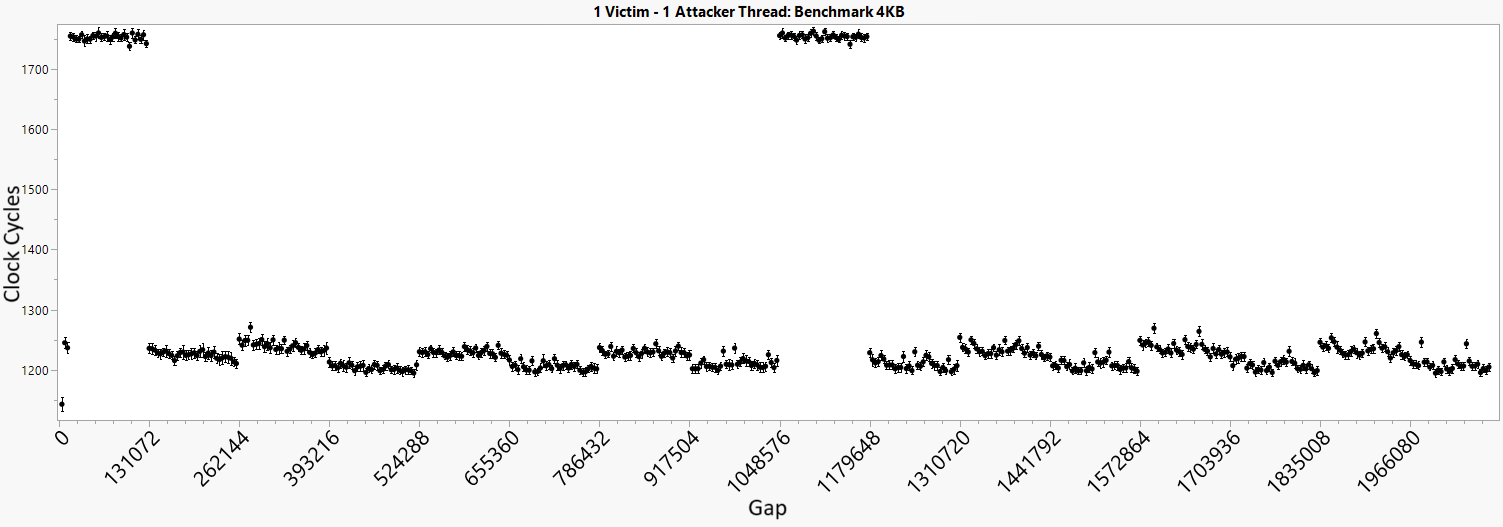}
    \caption{Workload 2MB / benchmark 4KB: 1 victim \& 1 attacker}
    \label{fig:4KB_512}
\end{figure*}

To force a new opening within the same chip where the victim's task is located, it is necessary to refer to Figure~\ref{fig:4KB_512}. In this case, we have changed the size of the workload, which is no more than 256KB, but we bring it to 2 MB. The individual benchmark of the victim task and the attacker is 4 KB. As explained in the \ref{design} Section, the attacker opens all row buffers in the same chip for the first three operations. From the fourth iteration onwards, the attacker navigates inside the same chip, once again forcing the DRAM to activate multiple row buffers at once, up to 128KB. From this moment on, until the attacker reaches 1 MB, he accesses all other chips. At the 256th iteration, the attacker reopens the chip where the victim is located just before rerunning the benchmark and creates row buffer contention for the following 128 KB. It is clearly visible again in Figure~\ref{fig:4KB_512} that the contention increases when we find the attacker in this condition because the clock cycles are identical to the initial part of the workload.


\subsection{Comparison}
Table \ref{tab:comparison} is a comparison between the data obtained from the Navigate, which does not have the purpose of attacking the critical task, and the Dram Bank\& Row-Conflict Bomb, which has the purpose of showing the contention obtained by analyzing the clock\_cycles of the victim critical task. 

\begin{table}[h]
\centering
\caption{Write experiment}
\begin{tabular}{|p{0.6cm}|p{0.7cm}|c|c|c|c|}
\hline
\textbf{Attac-kers} &\textbf{Bench-mark} & \textbf{Size} & \textbf{Navigate} & \textbf{RowConflict} & \textbf{+/- \%}\\
\hline
\hline
\textbf{\textit{1}} & \textbf{\textit{1KB}} & \textit{512 KB} & \textit{318212 ns} & \textit{509268 ns} & \textit{$\approx$ +60\%} \\
\hline
\hline

\textbf{\textit{2}} & \textbf{\textit{1KB}} & \textit{512 KB} & \textit{333410 ns} & \textit{506644 ns} & \textit{$\approx$ +51\%} \\
\hline
\hline

\textbf{\textit{3}} & \textbf{\textit{1KB}} & \textit{512 KB} & \textit{388034 ns} & \textit{555210 ns} & \textit{$\approx$ +43\%} \\
\hline
\hline

\textbf{\textit{7}} & \textbf{\textit{1KB}} & \textit{512 KB} & \textit{384698 ns} & \textit{970364 ns} & \textit{$\approx$ +150\%} \\
\hline
\hline

\end{tabular}
\label{tab:comparison}
\end{table}
As we can observe, using a 512KB workload and a 1KB benchmark as a reference, it is clear that with a number equal to 7 attacker threads, therefore filling all the positions in the RAM bank as in Figure \ref{fig:Row_Algorithm}, we have several clock\_cycles which increase by approximately 150\%. This shows that our initial assumptions were correct and that the result aligns with what was said.

\subsection{Comparision with Previous Work}
In this Section, we compare our approach with the algorithm employed by Farina et al. \cite{farinaISORC2022, giorgio, farina_thesis} and Bechtel et al. \cite{dos_mem_aware}. 

We selected the works by Farina et al. due to the similarities of processor and memory module architecture and frequency with that of our work. It should be noted that Farina et al. conducted experiments on an Intel Xeon Silver processor (Cascade Lake, 2.1 GHz, 8GB DDR4-2666 DRAM module), while we used an Intel Xeon Gold processor (Cascade Lake, 2.3 GHz, 8GB DDR4-2666 DRAM module). It is essential to recognize that the choice of processor and memory architecture plays a crucial role in these comparisons. The differences between Silver and Gold processors may impact the overall performance of the servers, but the differences are not significant with respect to our analysis, which is solely focused on stressing the DRAM and memory controller (and skipping the caches).

We consider the benchmark task under observation in experiments from Farina et al. as a victim and the co-executing stressing benchmarks as attacker threads.
Experiments by Farina et al. provided valuable data that allowed us to conduct some form of quantitative comparison, as shown in Table \ref{tab:comparison_with_farina}. Despite the small differences in the processor architecture and frequency, the variations in performance do not seem to deviate significantly from the data obtained using our Navigate algorithm for the purpose of comparison (especially when considering 2 and 3 attacker threads), as detailed in Table \ref{tab:comparison}.

\begin{table}[h]
\centering
\caption{Write experiment}
\begin{tabular}{|p{0.6cm}|p{0.7cm}|c|c|p{1.8cm}|c|}
\hline
\textbf{Attac-kers} &\textbf{Bench-mark} & \textbf{Size} & \textbf{Farina \cite{farina_thesis}} & \textbf{New Bank \& Row Conflict} & \textbf{+/- \%}\\
\hline
\hline
\textbf{\textit{1}} & \textbf{\textit{1KB}} & \textit{512 KB} & \textit{220968 ns} & \textit{509268 ns} & \textit{$\approx$ +130\%} \\
\hline
\hline

\textbf{\textit{2}} & \textbf{\textit{1KB}} & \textit{512 KB} & \textit{309222 ns} & \textit{506644 ns} & \textit{$\approx$ +63\%} \\
\hline
\hline

\textbf{\textit{3}} & \textbf{\textit{1KB}} & \textit{512 KB} & \textit{387916 ns} & \textit{555210 ns} & \textit{$\approx$ +43\%} \\
\hline
\hline

\textbf{\textit{7}} & \textbf{\textit{1KB}} & \textit{512 KB} & \textit{-} & \textit{970364 ns} & \textit{-} \\
\hline
\hline

\hline

\end{tabular}
\label{tab:comparison_with_farina}
\end{table}

Furthermore, it is essential to mention the data limitations encountered in the previous study limited us in conducting a thorough and comprehensive quantitative analysis. Since previous works did not present detailed experimental data considering a higher number of attacker threads, we were not able to compare the performance difference with 7 attacker threads. 

The approach by Bechtel et al. \cite{dos_mem_aware} serves as an excellent means to assess our methodical approach. Both our new approach and the chosen comparative study converge on the common goal of stressing internal hardware buffers, generating a series of cache misses that lead to DRAM congestion and interference in the DRAM bank of a victim task. Bechtel et al. strategically leverage mapping information from the underlying memory hardware, using Linux's HugePage support to control a portion of a physical address and influence the location of the DRAM bank during memory allocation. In our strategy, the Navigate algorithm (Section \ref{navigate}) provides a complete understanding of how DRAM manages requests and resources, allowing us to precisely target the victim task's DRAM bank and row without explicit knowledge of the memory mapping. Using the kmalloc() function facilitates efficient kernel memory acquisition. While the comparative work validates their attacks on Raspberry Pi 4 and Odroid XU4 using existing benchmarks from the SPEC CPU 2017 benchmark suite, our experimentation is based on the Intel Cascade Lake architecture and new stressing benchmarks (inspired by Farina et al. \cite{farinaISORC2022} and others from Section \ref{related_work}) targeted specifically towards the Intel Cascade Lake and our memory module architecture. Despite the divergence in methodologies, both approaches provide valuable insights into timing attacks via the memory, offering interesting directions for future work and future memory architectures.

Other works did not present detailed experimental data relevant to our work or used different underlying hardware architectures.
This absence of experimental data is contextualized by the fact that the main goal of the previous works was not exactly aligned with the objectives of our work.


\subsection{Conclusion}\label{sec:conclusion}
Understanding shared resources in multicore processors is essential to discovering all possible interference channels, which can help us regulate them adequately at runtime to provide timing guarantees for safety-critical applications despite co-executing unplanned or malicious applications. We focused on understanding commercial-of-the-shelf DDR4 DRAM dual in-line memory modules (DIMMs). We created the "navigate" algorithm to comprehend how DRAM manages requests and exploits parallelism to achieve higher performance. Based on this understanding, we explained how to create a timing attack ("DRAM Bank \& Row-conflict Bomb") on a victim task. We performed experiments on a Dell R640 server with 2nd generation Intel Xeon Gold processor (Cascade Lake) \cite{intel} \cite{uncore} using an 8GB DDR4-2666 ECC RDIMM (1Rx8) DRAM module and demonstrated that the DRAM Bank \& Row-Conflict Bomb significantly increases the observed execution time of the victim task (up to 150\%).

Future work involves creating a resource management algorithm for allocating DRAM memory locations to co-executing tasks at runtime to avoid the described timing attack, thus assisting in ensuring the safety and security of the critical applications. We also want to quantitatively compare our DRAM Bank \& Row-conflict Bomb to other similar benchmarks and timing attacks on different processor architectures. We expect that most use cases currently using DDR4 will migrate to DDR5 over time. Thus, we want to see the applicability of our approach and attack on DDR 5 memory modules.


\bibliography{bibliography}
\bibliographystyle{ieeetr}

\end{document}